\documentclass[prl,twocolumn,preprintnumbers,amsmath,amssymb]{revtex4}
\usepackage{amsfonts}
\usepackage{graphicx}
\usepackage{dcolumn}
\usepackage{float}
\usepackage{bm}

\begin{document}

%\righthyphenmin=1
%\emergencystretch=25pt

\title{An Improved Ehrenfest Approach to Model Correlated Electron-Nuclear Dynamics}

\author{Roman Baskov$^1$}
\author{Alexander White$^2$}
\author{Dmitry Mozyrsky$^2$}\email{mozyrsky@lanl.gov}
\affiliation{$^1$Institute of Physics of the National Academy of Sciences of Ukraine, pr. Nauky 46, Kyiv-28, MSP 03028, Ukraine
\\
$^2$Theoretical Division, Los Alamos National Laboratory, Los Alamos, NM 87545, USA}

\date{\today}

\begin{abstract}
Mixed quantum-classical mechanics descriptions are critical to modeling coupled electron-nuclear dynamics, \emph{i.e.} non-adiabatic molecular dynamics, relevant to photochemical and photophysical processes. We argue that, for polyatomic molecules, such mixed dynamics can not be \emph{efficiently} described in terms of a matrix gauge potential and develop the concept of a ``length gauge'' effective Hamiltonian, which helps clarifying certain aspects for the popular non-adiabatic computational approaches. In particular, within such an effective Hamiltonian formalism one readily derives the momentum rescaling boundary condition, used in the surface hopping algorithms. Furthermore, using the new formalism, we introduce a coupled Gaussian wavepacket parameterization of the nuclear wavefunction, which generalizes the Ehrenfest approach to account for electron-nuclei correlations. We test this new approach, Ehrenfest-Plus, on the standard set of model problems that probe electron-nuclear correlation in non-adiabatic transitions. The high accuracy of our approach, combined with mixed-quantum classical efficiency, opens a path for improved simulation of non-adiabatic molecular dynamics in realistic molecular systems.
  	
\end{abstract}

%\pacs{Need PACs}

\maketitle

\emph{Ab-initio} methods play an important role in the simulation of electronic and coupled electronic-nuclear processes in atomic, molecular, soft and condensed matter systems \cite{MARX2002,Basile18,Tavernelli15}. Light absorption and emission \cite{C5CP02100F,Guo15}, charge and energy transfer \cite{Jonas2018,Thiel16,Nelson14,Oberhofer17}, photoisomerization and photochemistry \cite{C5CP02100F,Mit08}, non-radiative relaxation \cite{C5CP00019J,Prezhdo17,Prezhdo14}, swift-ion stopping \cite{Correa12,Zeb12,Yost17}, \emph{etc.} are all influenced or controlled by the interaction of electronic excitation and nuclear motion. These \emph{non-adiabatic} processes are particularly difficult to model, due to the interaction of the quantum electrons and the nuclei. For the above processes, the two cannot be separated as in the traditional Born-Oppenheimer approximation. A fully quantum mechanical treatment of all the nuclei is computationally prohibitive for most systems\cite{Kendrick2018,Guo17,Guo16}. Mixed quantum-classical mechanics, which attempt to treat the nuclei semi-classically are therefore highly desirable \cite{Barbatti18,Barbatti11}.

One of the most successful \emph{first-principles} mixed quantum-classical algorithms is the mean-field Ehrenfest approach \cite{Li05,Kunisada16,Makhov17,Saita12}, in which the nuclei are subject to a classical force determined by the instantaneous average density of the electrons. However, as a mean-field approach, it cannot properly describe correlated electron-nuclear process. \emph{Ad-hoc} methods have been developed to treat these correlation effects, while maintaining computational efficiency \cite{Tully90}. Such methods, however, are unpredictable in terms of their accuracy, and no consensus exists on how to treat various situations \cite{Subotnik16,Nelson13,Barbatti18,Akimov14,Wang15,Wang14-2,Prezhdo97,Jaeger12,Martens16}.

Thus the development of \emph{ab-initio} non-adiabatic molecular dynamics (NAMD) approaches that closely tie to accessible mixed-quantum classical techniques can provide insight into these \emph{ad-hoc} algorithms, and potentially provide more reliable, but still numerically tractable, simulations.  Here we develop an approach that is specifically designed for \emph{on-the-fly} dynamics, typical to large scale NAMD. First we provide a general framework that naturally accounts for the fact that electronic properties are calculated for a specific nuclear configuration at a given time. From this generalization the mean-field Ehrenfest approach is derived, and then extended to account for fluctuations caused by electron-nuclear correlations. We call this extension the Ehrenfest-Plus method. We test this new approach on a set of standard models, which test electron-nuclear correlation in quantum scattering. However, we first begin by briefly reviewing the conventional approach to non-adiabatic processes in molecular systems.

A traditional description of  NAMD is based on the matrix vector potential picture, where the transitions between electronic states with energies $E_n({\bf x})$ parametrically dependent on $3N$-component coordinate vector ${\bf x}$ for the nuclear positions are described in terms of an effective  ``velocity gauge" Hamiltonian \cite{Baer06,Gherib16},
\begin{equation}\label{hamiltonian-gauge}
\hat{H}^{\rm vg} ({\bf x}) =\sum_{\mu=1}^{3N}{1\over 2M_\mu}{\big[\hat{p}_\mu-i\hat{{\cal A}}_\mu({\bf x})\big]^2}+\hat{E}(\bf x),
\end{equation}
where the non-adiabatic coupling vector (NACV), ${\cal A}_{nn^\prime,\mu}({\bf x}) = \langle n({\bf x}) |\partial_\mu {n^\prime}({\bf x})  \rangle$, and potential energy surface (PES) scalar, $E_{nn^\prime}({\bf x})=E_n({\bf x})\delta_{nn^\prime}$, potentials are matrices in the subspace spanned by the adiabatic electronic states $|n ({\bf x}) \rangle$ (eigenstates of the electronic Hamiltonian, $\hat{H}^e$). ${\hat p}_\mu$ is the conventional momenta operator acting in the ${\bf x}$-space, ${\hat p}_\mu = -i\partial_\mu$. $E_n({\bf x})$ is commonly referred to as the potential energy surface (PES) of state $n$.

%Unfortunately the description of the nuclear dynamics based on Eq. (\ref{hamiltonian-gauge}) is incompatible with MD simulations, where nuclei propagate along the (semi)classical trajectories and the basis functions $| n\rangle$ and the electronic energies $E_n$ (as well as their gradients) are evaluated only at the points along these trajectories. To be more specific,
One can arrive at Eq. (\ref{hamiltonian-gauge}) by considering elementary evolution of a full molecular state $|\Psi({\bf x})\rangle$. After a short time, $\epsilon$, this state evolves to $(1-i\epsilon {\hat H})|\Psi({\bf x})\rangle$, where $\hat H$ is the full molecular Hamiltonian, \emph{i.e.} the sum of the kinetic energy of the nuclei, $\hat{K}$, and the remaining ${\hat H}^e$ terms. Projecting the full molecular state onto the electronic subspace basis states, $\langle  n({\bf x}) |\Psi({\bf x})\rangle \equiv \psi_n({\bf x})$, the ionic states $\psi_n({\bf x})$ at time $\epsilon$ can be written as
\begin{align}\label{psi}
\psi_n({\bf x},\epsilon)= \sum_{n^\prime} \langle  n  ({\bf x})|\big[1-i\epsilon [{\hat K} + {\hat H}^e]\big]|{n^\prime}  ({\bf x}) \rangle\, \psi_{n^\prime}({\bf x},0)\,,
\end{align}
or equivalently (if $n$'s are eigenstates) the equation of motion for all $\psi$'s ($|\Psi ({\bf x})\rangle$) as
\begin{align}
~ i\partial_t|\Psi ({\bf x}) \rangle =\hat{H}^{\rm vg}({\bf x})|\Psi ({\bf x})\rangle\,;
\end{align}
see supplemental materials for details \cite{supp}. This equivalence requires that the eigenstates $|{n} ({\bf x}) \rangle$ are {\it globally} defined functions of the multi-dimensional position vector $\bf x$. %Indeed, matrix elements of $1$ and ${\hat H}^e$ in Eq. (\ref{psi}) are $\delta_{nn^\prime}$ and $E_n({\bf x})\delta_{nn^\prime}$, respectively, while the matrix element of ${\hat K}$ (recall that $\langle  n|{\hat K}|{n^\prime}\rangle$ is an operator in the ${\bf x}$-space), after simple manipulation \cite{baer} can be transformed into the first term of Eq. (\ref{hamiltonian-gauge}), so that in the limit $\epsilon\rightarrow 0$ we recover the Schrodinger equation $i\partial_t|\Psi\rangle =\hat{H}_{\rm gauge}|\Psi\rangle$.
However, for more than a few nuclear degrees of freedom, calculation of the PES and NACV matrices is numerically prohibitive. Thus \emph{on-the-fly} \emph{ab-initio} NAMD methods, where trajectories guide the calculation of the PES and NACV matrices, are desirable.

In these simulations, however, the basis states $|{n}\rangle$ are evaluated only {\it locally}, i.e. for a given position $\bar{\bf x}(t)$ along a trajectory. Since $\bar{\bf x}$ changes with time, the MD states $|{n (t) }\rangle$ are time-dependent (rather than ${\bf x}$-dependent, as we have assumed previously in the derivation of Eq. (\ref{hamiltonian-gauge})). %After infinitesimal time step $\epsilon$ they evolve to $|{n}\rangle +\epsilon \partial_t|{n}\rangle$.
%Therefore, when evaluating the matrix elements in Eq. (\ref{psi}) with respect to the time-dependent $|{n}\rangle$'s, we find that the matrix element of $1$ is $\delta_{nn^\prime}+\epsilon \langle  n|\partial_t{n^\prime}\rangle$ and
Thus an infinitesimal propogation ($t= t' +\epsilon$) of the wavefunction is given by
\begin{align}\nonumber
\label{psi2}
\psi_n({\bf x}, t)= \sum_{n^\prime}\big [ \delta_{nn^\prime}&+\epsilon \langle  n(t)|\partial_t{n^\prime(t')}\rangle
\\\nonumber &- i\epsilon\langle n(t)|{\hat H}|{n^\prime}(t')\rangle \big ] \psi_{n'}({\bf x},t')
\\
\langle n(t)|{\hat H}|{n^\prime}(t')\rangle \simeq &\sum_\mu {{\hat p}_\mu^2\over 2M_\mu} + \langle n(t)|{\hat H}^e({\bf x})| {n^\prime}(t')\rangle\,.
\end{align}
Note that since the matrix elements of $\hat H$ and the time derivative operator in Eq. (\ref{psi2}) are scaled by $\epsilon$, they must be evaluated to zero order in $\epsilon$. Thus we find that the molecular wavefunction $|\Psi\rangle$ in the basis of local electronic states $|{n}(t)\rangle$ satisfies the Schr{\"o}dinger equation with an effective ``length gauge" Hamiltonian,
\begin{equation}\label{hamlt-local}
{\hat H}^{\rm lg} = \sum_\mu {{\hat p}_\mu^2\over 2M_\mu} + \sum_{n,n^\prime} V_{nn^\prime}({\bf x},t) |{n^\prime}(t)\rangle\langle{n}(t)|\,,
\end{equation}
with
\begin{equation}\label{V}
V_{nn^\prime}({\bf x},t)= i \langle  n(t)|\partial_t{n^\prime}(t)\rangle + \langle n(t)|{\hat H}^e({\bf x})| {n^\prime}(t)\rangle\,.
\end{equation}

The potential energy in Eq. (\ref{V}) can be put in a more transparent form if we assume that the molecular wavefunction is sufficiently localized in the ${\bf x}$-space around a position $\bar{\bf x}(t)$. % (In this paper we choose ${x_0}$ to be the expectation value of ${\hat{x}}$ with respect to the current molecular state $|\Psi\rangle$.) Such assumption is natural when the nuclei are (semi)classical, i.e., have sufficiently high momenta and the potentials $V_{nn^\prime}({\bf x})$ are slowly varying functions of ${\bf x}$.
Then we expand ${\hat H}^e$ as
\begin{equation}\label{He}
{\hat H}^e({\bf x})={\hat H}^e(\bar{\bf x}) + [\partial{\hat H}^e(\bar{\bf x})/\partial \bar{\bf x}]\cdot[{\bf x}-\bar{\bf x}]+...\,.
\end{equation}
Then, choosing $|{n(t)}\rangle$ to be a local adiabatic basis (i.e. at point $\bar{\bf x}$(t)) we can readily evaluate the matrix elements in Eq. (\ref{V}). For $n=n^\prime$ the second matrix element in Eq (\ref{V}) gives $E_n(\bar{\bf x}) - {\bf f}_n(\bar{\bf x})({\bf x}-\bar{\bf x})$, where ${\bf f}_n$ is the classical electron-nucleus force for the $n$'s  PES, $E_n(\bar{\bf x})$. For $n\neq n^\prime$, by virtue of Hellmann-Feynman theorem, this matrix element is $\Delta E_{nn^\prime}(\bar{\bf x}){\cal{\bf {\cal A}}}_{nn^\prime}(\bar{\bf x})({\bf x}-\bar{\bf x})$, where ${\cal{\bf {\cal A}}}_{nn^\prime}$ is the  $n,n^\prime$ element of the NACV matrix (Eq. \ref{hamiltonian-gauge}) and $\Delta E_{nn^\prime}=E_n-E_{n^\prime}$. Furthermore, in the absence of magnetic field, the electronic Hamiltonian ${\hat H}^e$ is real and therefore the states $|{n} (t)\rangle$ can also be chosen real. Then, for diagonal $n=n^\prime$ terms, the first matrix element in Eq. (\ref{V}) vanishes, while the off-diagonal   $V_{nn^\prime}$ can be written as
\begin{equation}\label{Voff}
i {\cal{\bf {\cal A}}}_{nn^\prime}\cdot\bar{\bf v} + \Delta E_{nn^\prime} {\bf {\cal A}}_{nn^\prime} \cdot ({\bf x}-\bar{\bf x})
\simeq i {\bf {\cal A}}_{nn^\prime}\cdot\bar{\bf v} \,e^{i\Delta {\bf p}_{nn^\prime}\cdot({\bf x}-\bar{\bf x})},
\end{equation}
where $\Delta{\bf p}_{nn^\prime}= -\Delta E_{nn^\prime} {\bf{\cal A}}_{nn^\prime}/({\bf{\cal A}}_{nn^\prime}\cdot\bar{\bf v} )$ and $\bar{\bf v} \equiv {\dot{\bar {\bf x}}}$. The second equality in Eq. ({\ref{Voff}}) implies that $i\Delta {\bf p}_{nn^\prime}\cdot({\bf x}-\bar{\bf x})\ll 1$, which is the case only if the molecular state $|\Psi\rangle$ is sufficiently localized around position $\bar{\bf x}$, as we have already assumed in Eq. (\ref{He}).

Thus Eq. (\ref{hamlt-local}) in the ``local adiabatic" basis can be written as
\begin{eqnarray}\label{hamlt_local_adiab}
\hat{H}^{\rm la}\simeq \sum_\mu {{\hat p}_\mu^2\over 2M_\mu}  + \sum_{n}\big[E_n(\bar{\bf x}) - {\bf f}_n(\bar{\bf x})\cdot({\bf x}-\bar{\bf x}) \big]|n\rangle\langle n|\nonumber\\
+\sum_{\substack{n\neq n'}}  i {\cal{\bf {\cal A}}}_{nn^\prime}(\bar{\bf x})\cdot\bar{\bf v} \,e^{i\Delta {\bf p}_{nn^\prime}\cdot({\bf x}-\bar{\bf x})}        |n\rangle\langle n'|\,,~~~~
\end{eqnarray}
where we have used a shorthand notation $|n\rangle \equiv | n(t)\rangle$. The physical significance of the phases in the off-diagonal coupling coefficients in Eq. (\ref{hamlt_local_adiab}) can be readily understood if we assume that, say, at time $t=0$ the wavefunction is given by $|\Psi(0)\rangle = g(\bar{\bf x},\bar{ \bf p},{\bf x},0) |n_0\rangle$,
\begin{equation}\label{gaussian}
g(\bar{\bf x},\bar{ \bf p}, {\bf x},t) = {\cal N} e^{i[{\bf x}-\bar{\bf x}(t)]\cdot \hat{\alpha}(t) \cdot[{\bf x}-\bar{\bf x}(t)]+i\bar{\bf p}(t)\cdot[{\bf x}-\bar{\bf x}(t)]}\,,
\end{equation}
where $\hat{\alpha}$ is a complex matrix, $\alpha_{\mu\nu}= \alpha_{\mu\nu}^{\rm Re}+i\alpha_{\mu\nu}^{\rm Im}$ and ${\cal N}=[2^{3N}{\rm \det}(\hat{\alpha}^{\Im})/\pi^{3N}]^{1/4}$ is the normalization coefficient. The Gaussian in Eq. (\ref{gaussian}) describes a nuclear subsystem centered around classical position $\bar{\bf x}$ and having momentum $\bar{\bf p}$. Upon application of the off-diagonal interaction term in Eq. ({\ref{hamlt_local_adiab}}), the electronic state $|n'\rangle$ switches to state $|n\rangle$, while the initial momentum $\bar{\bf p}$ changes to $\bar{\bf p}+\Delta {\bf p}_{nn'}$, and the phase changes accordingly. The momenta of the old and the new wavepackets approximately satisfy classical energy conservation (\emph{i.e.} for ${\bf \bar p}\gg\Delta {\bf \bar p}_{nn'}$). Such a prescription for momentum rescaling has been utilized in numerous \emph{ad-hoc} numerical approaches, such as surface hopping. We emphasize that the ``energy conservation" is a direct consequence of the choice of adiabatic basis set $| n(t)\rangle$ in Eq. (\ref{hamlt_local_adiab}). If $| n(t)\rangle$ is not to be the eigenstates of ${\hat H}^e(\bar{\bf x}(t))$, the off-diagonal couplings in Eqs. (\ref{Voff}) and (\ref{hamlt_local_adiab}) do not have the exponential form with phases $\Delta {\bf p}_{nn^\prime}\cdot({\bf x}-\bar{\bf x})$.

The choice of $\vert n(t) \rangle$'s in Eq. (\ref{hamlt-local}) is not limited to the local adiabatic eigenstates of ${\hat H}^e(\bar{\bf x}(t))$. Instead, the basis functions can be defined as linear combinations of these adiabatic states,
\begin{equation}\label{Ehr-states}
| m(t)\rangle=\sum\limits_{n} c_{mn}(t)| n(t)\rangle\,,
\end{equation}
with coefficients $c_{m,m'}(t)$ chosen to ensure the orthogonality of states $| m(t)\rangle$ at any given time $t$. The coupling coefficients $V_{mm'}({\bf x},t)$ will still have similar form to Eq. (\ref{V}). Since we have a freedom in choosing the coefficients in Eq. (\ref{Ehr-states}), we may require that these coefficients are chosen to ensure the condition
\begin{equation}\label{Ehr-cond}
V_{mm^\prime}(\bar{\bf x})=0
\end{equation}
for $m\neq m^\prime$. Then, if the molecular state is well localized around $\bar{\bf x}$, one can argue that the corrections due to second term in the rhs of Eq. (\ref{He}) can be neglected. Therefore, within such an approximation, in the basis of $| m(t)\rangle$'s, the transitions between the states with different $m$'s are absent, and, the molecular wavefunction has a single component ($m_0$) in the time-dependent electronic state basis.

With the use of Eqs. (\ref{V}) and (\ref{Ehr-states}), condition (\ref{Ehr-cond}), after a straightforward manipulation, can be rewritten as
\begin{equation}\nonumber
i\dot{c}_{m_0n}(t)=E_n(\bar{\bf x})c_{m_0n}(t)+i\sum_{n^\prime}{\bf {\cal A}}_{nn^\prime}(\bar{\bf x})\cdot{\bf \bar v}\,c_{m_0n'}(t)\,.
\end{equation}
Furthermore the force that acts on the nuclei in state $| m_0\rangle$, i.e., $- \langle m(t)| \partial_{\bar{\bf x}}{\hat H}^e(\bar{\bf x}) | m(t)\rangle$, can be written as
\begin{equation}\nonumber
{\bf f}_{m_0} (t) = -\sum_{n,n^\prime} \langle {n} (t) | \partial_{\bar{\bf x}}{\hat H}^e(\bar{\bf x}) |{n^\prime}(t)\rangle c_{{m_0}n}^\ast(t)c_{{m_0}n'}(t)    \,.
\end{equation}
Thus, we see that the condition (\ref{Ehr-cond}) corresponds to Ehrenfest dynamics \cite{Li05}, where the average nuclear postion, $\bf \bar x$, propagate along a single, average, PES. The main shortcoming of the method is related to the neglect of nuclear fluctuations associated with $(\partial{\hat H}^e/\partial \bar{\bf x})\cdot({\bf x}-\bar{\bf x})$ term in Eq. (\ref{He}) . Such fluctuations are described, for example, by the phases in the off-diagonal matrix elements in the Hamiltonian in Eq. (\ref{hamlt_local_adiab}). As we have discussed above, these phases lead to the difference in momenta of the wavepackets moving along different adiabatic PES, which is not the case for the Ehrenfest method.

The surface hopping method \cite{Tully90}, on the other hand, does account for the phases in the couplings in Eq. (\ref{hamlt_local_adiab}) by adjusting the momenta of the wavepacket after it ``hops" between PESs. Yet, the \emph{ad-hoc} Markovian assumption for the hopping rate and the lack of quantum interference between trajectories may lead to significant uncontrollable errors \cite{Tully90}.

{\it Computational approach:} To simplify, we will assume that there are only two relevant time-dependent locally-adiabatic electronic states, $|1(t)\rangle$ and $|2(t)\rangle$. Generalization to a higher number of electronic states is straightforward. The nuclear wavefunction has two-components and the corresponding Schr{\"o}dinger equation, using Eq. (\ref{hamlt_local_adiab}), reads
\begin{align}\label{init_eq_sys}
%                  i\dot{\psi_1}({\bf x},t)=\hat{h}_1\psi_1({\bf x},t)+V_{12}({\bf x})\psi_2({\bf x},t)
%                 \\\nonumber
%                 i\dot{\psi_2}({\bf x},t)=\hat{h}_2\psi_2({\bf x},t)+V_{21}({\bf x})\psi_1({\bf x},t)
                  i\dot{\psi_1}({\bf x},t)=H^{\rm la}_{11}(\bar{\bf x})\psi_1({\bf x},t)+H^{\rm la}_{12}(\bar{\bf x})\psi_2({\bf x},t)
                  \\\nonumber
                  i\dot{\psi_2}({\bf x},t)=H^{\rm la}_{22}(\bar{\bf x})\psi_2({\bf x},t)+H^{\rm la}_{21}(\bar{\bf x})\psi_1({\bf x},t)
\end{align}
%with
%\begin{equation}\label{h1}
%{\hat{h}}_n= \sum_\mu {{\hat p}_\mu^2\over 2M_\mu}  + E_n(\bar{\bf x}) - {\bf f}_n(\bar{\bf x})({\bf x}-\bar{\bf x})
%\end{equation}
%and $V_{12}({\bf x})=i{\bf {\cal A}}_{12}(\bar{\bf x})\cdot{\bf v}_0\,e^{i\Delta {\bf p}_{12}({\bf x}-\bar{\bf x})}$, $V_{12}=V^\ast_{21}$.
We take $\bar{\bf x}(t)$ to be the mean coordinate,
\begin{equation}\nonumber
\bar{\bf x}(t)=\int d{\bf x}~(\psi^*_1{\bf x}\psi_1 + \psi^*_2{\bf x}\psi_2)\,,
\end{equation}
and $\bar{\bf v}(t)$ to be the mean velocity,
\begin{equation}\nonumber
\bar{v}_{\mu}(t)=\int d{\bf x} ~(\psi^*_1{\hat p}_\mu\psi_1 + \psi^*_2{\hat p}_\mu\psi_2)\hat{M}^{-1}_{\mu}\,.
\end{equation}

We assume that initially the system is a local adiabatic state, $\vert \Psi (0) \rangle = \psi_1({\bf x},0)|1(0) \rangle$, with nuclear state $\psi_1$ being a Gaussian, \emph{i.e.} $\psi_1({\bf x},0) = g({\bf x}_1,{ \bf p}_1,{\bf x},0)$ from Eq. (\ref{gaussian}).
% $\bar{\bf x} \equiv {\bf x}_1(0)$, $\bar{\bf p} \equiv {\bf p}_1(0)$ and a complex matrix $\alpha_{\mu\nu}(0)$ defining the wavepacket's widths in coordinate and momentum spaces.
Note that, more generally, any initial $\psi({\bf x},0)$ can be represented by a sum of Gaussians.

In the absence of coupling, ${\cal A}_{12}=0$, $\psi_1$ retains it's Gaussian form during propagation by Eq. \ref{init_eq_sys} with Eq. (\ref{hamlt_local_adiab}). The Gaussian coefficients satisfy equations of motions \cite{Heller75}:
\begin{align}\nonumber
{\dot x}_{1\mu} &\equiv \dot{x}_{0\mu}(t) = {\dot p}_{1\mu}/M_\mu,
\\\nonumber {\dot{\bf p}}_1 &= {\bf f}_1({\bf x}_1),
\\\nonumber{\dot \alpha}_{\mu\nu} &= -2\sum_{\lambda}\alpha_{\mu\lambda}\alpha_{\lambda\nu}/M_\lambda\, .
\end{align}

%The gaussian approximation, while formally exact for linear and q3uadratic potentials, such as in Eq. (\ref{init_eq_sys}), typically breaks down at long times if the nonlinearities, i.e., the terms denoted by $...$ in Eq. (\ref{He}), are sufficiently strong. However, the non-adiabatic regions are typically quite narrow and so the we are interested in sufficiently short time dynamics when the effects due to the nonlinearities have not yet developed.

In the presence of coupling, ${\cal A}_{12}\neq0$ we expect that, for short times, %the Gaussian shape of functions $\psi_1$ and $\psi_2$ is not retained. This is due to separation of $\psi_1$ and $H_{\rm la,12} \psi_2$ (or $\psi_2$ and $H_{\rm la,21} \psi_1$) in phase space ($\bar{\bf x} \neq {\bf x}_1 \neq {\bf x}_2$).
%However, if $H_{\rm lg,12}\psi_2({\bf x})$ is similar to $\psi_1({\bf x})$, we expect the Gaussian anzatz to be a good solution to the equation for $\psi_1$. Indeed, in such case the last term in Eq. (\ref{init_eq_sys}) (for $\psi_1$) just merely redefines $E_1(\bar{\bf x})$ in this equation. Furthermore, due to the fact that $V_{12}({\bf x})V_{21}({\bf x})$ is ${\bf x}$-independent, and that $V_{21}({\bf x})\psi_2({\bf x})$ is also a gaussian (provided that $\psi_2({\bf x})$ is a gaussian wavepacket), we can expect the gaussian anzatz
\begin{align}\label{anzatz}
\psi_1(t) = c_1(t)\,g_1({\bf x}),
\\\nonumber  \psi_2(t) = c_2(t)\,g_2({\bf x})\,,
\end{align}
($\,g_n({\bf x}) \equiv g({\bf x}_n,\,{\bf p}_n,\,{\bf x})\,$) is a good solution, provided that
\begin{equation}\label{cond}
g_1(\,{\bf x})\propto e^{i\Delta {\bf p}_{12}\cdot({\bf x}-\bar{\bf x})}g_2({\bf x}),
\end{equation}
during the course of evolution. Condition (\ref{cond}) breaks down when the wavepackets in Eq. (\ref{anzatz}) spatially separate due to the difference in forces ${\bf f}_1$ and ${\bf f}_2$ in Eq. (\ref{hamlt_local_adiab}). However, if the wavepackets traverse the non-adiabatic regions rapidly, the condition (\ref{cond}) holds approximately while the time-dependent coupling ${\bf {\cal A}}_{12}(\bar{\bf x})\cdot{\bf \bar v}$ is non-zero. %and, as confirmed by our simulations (see below), the anzatz typically holds until the wavepackets are out of the non-adiabatic region.

To find the coefficients $c_1$ and $c_2$ we project Eq. (\ref{init_eq_sys}) for $\psi_1$ onto state $\psi_2$ and vice versa. Then we find that %
%\begin{equation}\label{coeff_eqs}
%    \left\{
%                \begin{array}{ll}
                  %\dot{c}_1=-\langle g_1|\frac{\partial g_1}{\partial t}\rangle c_1{-}i\langle g_1|\hat{h}_1|g_1\rangle c_1{-}i\langle g_1|V|g_2\rangle c_2\\
                  %\dot{c}_2=-\langle g_2|\frac{\partial g_2}{\partial t}\rangle c_2{-}i\langle g_2|\hat{h}_2|g_2\rangle c_2{-}i\langle g_2|V^\ast|g_1\rangle c_1\,
%                \end{array}
 %             \right.
%\end{equation}\label{coeff_eqs}
\begin{align}\label{coeff_eqs}
\dot{c}_n(t)=-\int d{\bf x} ~&\Big[g_n^\ast({\bf x}) {\dot g_n({\bf x})} c_n(t) \\\nonumber
&{-}i\sum_{n'} g_n^\ast({\bf x}) H^{\rm la}_{nn'}({\bf x},\bar{\bf x}) {g_{n'}({\bf x})} c_{n'}(t)\Big]\,.
\end{align}
%where the matrix elements $\langle g_1|g_2\rangle$, etc., are defined as $\int d{\bf x}\, g_1^\ast({\bf x})g_2({\bf x})$, etc.
Since $g_1$ and $g_2$ are Gaussian functions, e.g. Eq. (\ref{gaussian}), the matrix elements can be calculated analytically; the explicit expressions are presented in the Supplemental Materials \cite{supp}.

Equations of motion for the parameter ${\bf x}_n(t)$  and ${\bf p}_n(t)$ and $\alpha_{n,\mu\nu}(t)$ defining the Gaussians can be found by relating these quantities to the expectation values of various operators. ${\bf x}_n(t)=\int d{\bf x}~{\psi^*_n{\bf x}\psi_n/{\vert c_n\vert^2}}$ and ${\bf p}_n(t)=\int d{\bf x}~{\psi^*_n{\bf \hat{p}}\psi_n/{\vert c_n\vert^2}}$ are the state dependent expectation values of coordinate and momentum operators. Using Eqs. (\ref{init_eq_sys}) and (\ref{anzatz}), after some algebra, one arrives at
\begin{subequations}\label{param_ev_eqn}
\begin{align}
%\dot{\bf x}_{1}&={\bf v}_{1}-\frac{2{\rm Im}\{c_1c^\ast_2\langle g_2|V^\ast\cdot (\hat{x}_\mu-x_{1\mu})|g_1\rangle\}}{|c_1|^2}\,,\\
%\dot{\bf p}_{1}&=f_{1\mu}(\bar{\bf x})-\frac{2{\rm Im}\{c_1c^\ast_2\langle g_2|V^\ast\cdot (\hat{p}_\mu-p_{1\nu})|g_1\rangle\}}{|c_1|^2}\,,
\dot{\bf x}_{n}&={\bf v}_{n}
\\\nonumber&-\sum_{n'\neq n}2{\rm Im}\{\frac{c_nc^\ast_{n'}}{|c_n|^2}\int d{\bf x} g_{n'}^\ast({\bf x})  H^{\rm la}_{n'n}({\bf \bar x}) [{\bf x}-{\bf x}_{n}]g_n({\bf x})\}\,,\\
\dot{\bf p}_{n}&={\bf f}_{n}(\bar{\bf x})
\\\nonumber&-\sum_{n'\neq n}2{\rm Im}\{\frac{c_nc^\ast_{n'}}{|c_n|^2}\int d{\bf x} g_{n'}^\ast({\bf x})  H^{\rm la}_{n'n}({\bf \bar x}) [{\bf \hat p}-{\bf \hat p}_{n}]g_n({\bf x})\}\,,
\end{align}
\end{subequations}
with $n'\neq n$. Furthermore, the quantities $\alpha_{n\mu\nu}(t)$ can be expressed in terms of the expectation values of the products of coordinate and momentum operators. Specifically, elements of the inverse matrix ${\bf\alpha}^{-1}$ is related to the square deviations of ${\bf x}$ as
\begin{equation}\nonumber
\int d{\bf x}  ~g_n({\bf x}) [{\bf x}-{\bf x}_n]\otimes [{\bf x}-{\bf x}_n] g_n({\bf x}) = \frac{1}{4} [{\bf \hat \alpha}_n^{\rm Im}]^{-1} ,
\end{equation}
and, similarly,
\begin{equation}\nonumber
\int d{\bf x}  ~g_n({\bf x}) [{\bf x}-{\bf x}_n]\otimes [{\bf \hat p}-{\bf p}_n] g_n({\bf x}) = {\bf \hat \alpha}_n^{\rm Re}\cdot[{\bf \hat \alpha}_n^{\rm Im}]^{-1} .
%\langle g_n|(x_{\mu}-x_{n\mu})({\hat p}_\nu-p_{n\nu})|g_n\rangle= [{\bf\alpha}_n^{\rm Re}\cdot({\bf \alpha}_n^{\rm Im})^{-1}]_{\mu\nu}\,.
\end{equation}
Then, after some algebra we obtain
\begin{align}\label{eq-alpha}
\nonumber&\dot{\hat \alpha}_n={-}2\hat \alpha_n\cdot {\hat M}^{-1}\cdot \hat \alpha_n {-}\sum_{n'\neq n}\Big\{ \frac{2c^\ast_nc_{n'}}{|c_n|^2} \int d{\bf x} ~ g^\ast_n({\bf x}) H^{\rm la}_{n,n'} ({\bf \bar x})
\\&\times\{4\hat \alpha_n^{\rm Im}\cdot[{\bf x}{-}{\bf x}_n]\otimes  [{\bf x}{-}{\bf x}_n]\cdot \hat \alpha_n^{\rm Im} {-}\hat\alpha_n^{\rm Im} \} g_{n'}({\bf x})\Big\}.
\end{align}
Eqs. (\ref{coeff_eqs}, \ref{param_ev_eqn}) and (\ref{eq-alpha}) complete the approximate evolution of the wavefunction according to the Hamiltonian in Eq. (\ref{init_eq_sys}). While individual nuclear configurations, ${\bf x}_n$'s, are propagated, electronic structure calculations are only required for the ${\bf \bar x}$ nuclear configuration, similar to traditional Ehrenfest method. This is a significant departure from traditional Gaussian wavepacket methods, which calculate electronic structure at the center of each Gaussian  \cite{AIMS00,AIMC14,Richings15,White16,White15,White14,Meek16,Vacher2016,C6FD00073H}. Thus, we call this new approach Ehrenfest-Plus (EP), as it incorporates the equation of motion for the state dependent Gaussian variables in addition to the mean-field variables.
\begin{figure*}[ht!]
\includegraphics[width=0.99\textwidth,keepaspectratio]{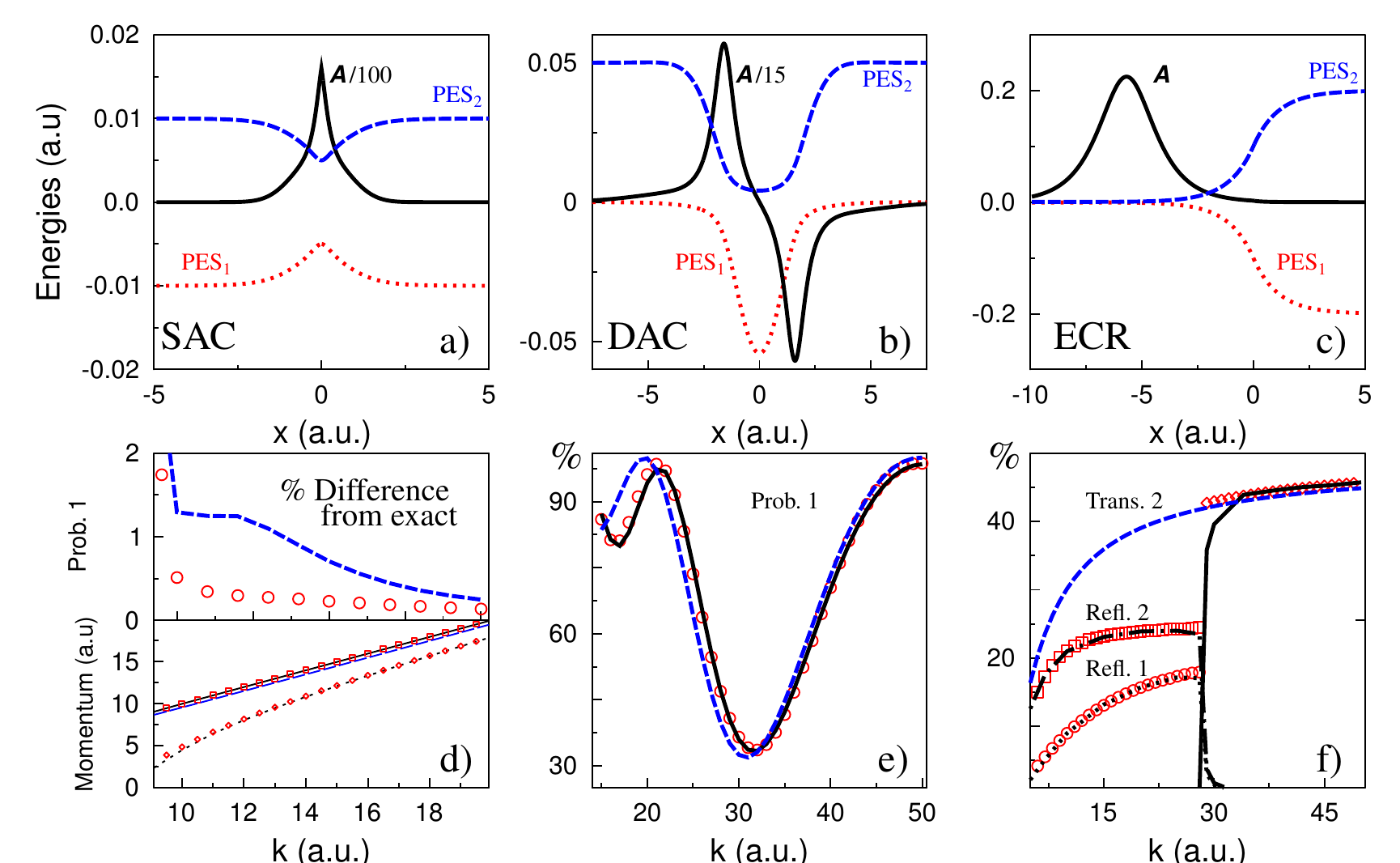}			
\caption{\label{Results} a-c) Model PESs for state 1 (red dot) and 2 (Blue dashed) with scaled non-adiabatic coupling vector (NACV) between states 1 and 2 (black solid). a) SAC, b) DAC, c) and ECR potentials. d) SAC results: {\it Upper panel}: difference between the exact scattering probability on surface 1 and the Ehrenfest (blue-dashed) or EP (red circles) vs. incoming momentum ($k$). {\it Lower panel}: outgoing momenta, exact on surface [1/2] (black [solid/dashed] line ), EP on surface [1/2] (red [squares/diamonds]), and Ehrenfest (blue dashed line) vs $k$. e) Probability of Transmission on surface 1 for the DAC model, exact (black line), Ehrenfest (blue dashed), and EP(red circle). f) ECR scattering probabilities, Ehrenfest probability for state 2 (blue dashed line), [exact/EP] transmission probability on state 2 [black solid line/red diamonds], [exact/EP] Reflection probability on state 2 [black dash-dot line/red squares], and  [exact/EP] Reflection probability on state 1 [black dotted line/red circles].     }

\end{figure*}
Note that since we assume that $c_1(0)=1$ and $c_2(0)=0$, in order to avoid divergence in the second terms in the rhs of Eqs. \ref{param_ev_eqn} and \ref{eq-alpha}, one needs to choose the initial parameters by
\begin{align}\label{in-cond}
{\bf x}_2(0)&={\bf x}_1(0)\,,
\\\nonumber{\bf p}_2(0)&={\bf p}_1(0)+\Delta{\bf p}_{12}(0)\,,
\\\nonumber { \hat \alpha}_2(0)&={\hat  \alpha}_1(0).
\end{align}
Eqs. (\ref{in-cond}) uniquely determine the initial values of the second wavepacket (with zero amplitude at $t=0$) by guaranteeing that the second terms in Eqs. (\ref{param_ev_eqn}) and  (\ref{eq-alpha}) are finite.

The approximations presented are only valid for a finite time.  As the wavepackets separate in phase space, ($\bar{\bf x} \neq {\bf x}_1 \neq {\bf x}_2$ \text{and ${\bf p}_1 \neq {\bf p}_2 - \Delta {\bf p}_{12}$}), the two wavepacket picture will become incomplete. The equations of motion of the wavepackets must ``separate", allowing new wavepackets to be ``spawned", similar to the \emph{ab-initio} multiple spawning \cite{AIMS00} or decoherence induced surface hopping \cite{Jaeger12} approaches. Thus, at the spawning point we begin to propagate multiple and independent sets of Eq. (\ref{init_eq_sys}). This effectively resets the variables, ($\bar{\bf x} = {\bf x}_1 = {\bf x}_2$ \text{ and ${\bf p}_1 = {\bf p}_2 - \Delta {\bf p}_{12}$}), at the cost of propagating an additional set of wavepackets.  While reducing the time between ``spawns" can control the accuracy, this will exponentially increase the number of simulations.
The natural spawning criteria are:
\begin{align}\label{criterion}
\vert c_n(t\neq 0) \vert &\le c_{min}\,,
\\
\vert {\cal A}_{mn}({\bf \bar x})\cdot {\bf \bar v} \vert &\le {\cal A}v_{min}\,,
\\
\vert \int d{\bf x} ~g_m^\ast({\bf x}) e^{i\Delta {\bf p}_{mn}\cdot({\bf x}-\bar{\bf x})} { g_n({\bf x})}  \vert &\le O_{min}\,.
\end{align}
The first criterion results from the divergence in Eqs. (\ref{param_ev_eqn}) and (\ref{eq-alpha}) when $|c_n|^2=0$. However, ``spawning" at this point does not increase the number of wavepackets as there is no need to propagate a wavepacket with no amplitude. The second criterion comes from the divergence of $\Delta {\bf p}_{mn}$ when $A_{mn}({\bf \bar x})\cdot {\bf \bar v}=0$. The phase of the coupling term, Eq. (\ref{hamlt_local_adiab}), changes rapidly at this point, quickly invalidating approximation (\ref{anzatz}). The third criterion is a direct measure of approximation (\ref{anzatz}). While taking $O_{min}$  close to one will negate the need for the other criterions, it can create an undesirably high ``spawning" rate \cite{White16}. In practice, for the following simulation results, we are using only the second criterion: The spawning occurs only when $A_{mn}({\bf \bar x})\cdot {\bf \bar v}$ changes sign.

{\it Simulation results:}
We compare our EP results to the exact Schr{\"o}dinger equation and Ehrenfest method for three model problems, which are standard tests of electron-nuclear correlation in non-adiabatic transitions, see Figure. \ref{Results}. In these three models a wavepacket with initial  $p_1(0)=k$, $x_1(0)=-15$, and $\alpha^{\text{Im}}(0)=k^2/400$ on the first PES, $c_1(0)=1$ and $c_2(0)=0$ is propagated through the region of finite NACV.

The first model is a single avoided crossing (SAC). Here, Ehrenfest and the Ehrenfest-Plus both quantitatively agree with the exact solution for scattering probabilities, with a slight increase in accuracy for EP. However the EP method correctly calculates the outgoing momenta of the wavepackets on the two surfaces, while the Ehrenfest method results in a single ``average'' momentum.

For the second model, a double avoided crossing (DAC), quantum interference between two pathways (crossing $x=0$ on either surface 1 or 2) leads to ``Stueckelberg" oscillations in the scattering probability. By neglecting difference in forces on PES 1 and 2, the Ehrenfest method shifts the phase of the oscillations at low $k$, whereas the EP method correctly captures this interference, with a single branching at the point where the NACV changes sign.

The third model includes an extended coupling region and a reflection (ECR) on PES 2 for $k$ less than $\sim30$. For $k<30$ after initial crossing of the finite NACV region, the wavepacket on PES 2 will reflect and re-enter the NACV region. This will cause probability to be transfered from state 2 back to state 1. Unlike the DAC model, that pathway (leading to reflection on PES 1) does not interfere with the wavepacke that transmitted on PES 1 after the first crossing.  This is accurately represented in the exact solution and EP method. However, by only propagating an ``average" momentum, the Ehrenfest method misses the reflection entirely. As in the DAC model, the EP method requires a single branching at the reflection point on PES 1.

In summary we have developed a general approach, based on expansion of the molecular wavefunction into the basis of ``local", time-dependent, electronic wavefunctions, \emph{i.e.} dependent on the nuclear positions at a particular time. The commonly used Ehrenfest method is re-derived from this general approach. In contrast, we use the local eigenstates of the electronic Hamiltonian, defined at the \emph{mean positions}, to build a new Gaussain propagation scheme. This approach directly leads to well defined, \emph{first principles}, boundary conditions for a ``spawning/hopping" scheme, and new ``beyond classical" equations of motion for the Gaussian variables. We apply the method on standard model problems that illustrate the effects of electron-nuclear correlation on non-adiabatic transitions.  Our method quantitatively reproduces the exact Shr{\"o}dinger equation results in these models, to within a few percent, for both scattering probabilities and momenta.
Due to numerical efficiency that is similar to Ehrenfest dynamics, it is feasible to apply this method to realistic molecular species.

\begin{acknowledgments}We acknowledge support of the U.S. Department of Energy through the Los Alamos National Laboratory (LANL) LDRD Program. LANL is operated by Los Alamos National Security, LLC, for the National Nuclear Security Administration of the U.S. Department of Energy under Contract No. DE-AC52- 06NA25396. We also acknowledge the LANL Institutional Computing (IC) Program provided computational resources.
\end{acknowledgments}

\newpage

\widetext

\appendix
\section{Supplemental Materials}

\subsection{``Velocity Gauge" Hamiltonian, Eqs. 1 and 3}
Starting from main text Eq. 2.
\begin{align}
\psi_n({\bf x},\varepsilon)= \sum_{n^\prime} \langle  n  ({\bf x})|\big[1-i\varepsilon [{\hat K} + {\hat H}^e]\big]|{n^\prime}  ({\bf x}) \rangle\, \psi_{n^\prime}({\bf x},0)\,,
\end{align}
$\vert  n  ({\bf x}) \rangle$ is taken as the eigenvectors of $\hat{H}^e$ which depend on the positions of the nuclei ${\bf x}$. Thus $\langle{n}|\hat{H}^e|{n'}\rangle \equiv E_n \delta_{nn'}$. the nuclear kinetic energy operator is given by ${\hat K} = \frac{-1}{2m}\frac{\partial}{\partial {\bf x}}\cdot \frac{\partial}{\partial {\bf x}} \equiv \frac{-1}{2m}\frac{\partial^2}{\partial {\bf x}^2}$. We assume all nuclei have the same mass, $m$, to simplify notation, but in general it is a diagonal matrix.

\begin{align}
\\\nonumber
 \langle n ({\bf x})|[1-i\varepsilon [{\hat K} + {\hat H}^e]\big]|{n' ({\bf x})}\rangle=\delta_{nn'}{-}i\varepsilon\bigg[-\frac{\langle n({\bf x})|\frac{\partial^2}{\partial {\bf x}^2}|{n'({\bf x})}\rangle}{2m}{+}E_{n}(x)\delta_{nn'}\bigg]
 \\\nonumber
- \langle n ({\bf x})|\frac{\partial^2}{\partial {\bf x}^2}|{n'({\bf x})}\rangle=-\langle n ({\bf x})|\frac{\partial^2{n'({\bf x})}}{\partial {\bf x}^2}\rangle-i2{\cal A}_{nn'}({\bf x})\cdot {\bf \hat p}+\delta_{nn'}{\bf \hat p}^2,
 \end{align}
 where  ${\cal A}_{nn^\prime,\mu}({\bf x}) \equiv \langle n({\bf x}) |\partial_\mu {n^\prime}({\bf x})  \rangle$, and $-i\frac{\partial}{\partial {\bf x}} \equiv \hat{\bf p} $. Since $[\frac{\partial}{\partial {\bf x}} \langle n ({\bf x})\vert n'({\bf x})\rangle] \equiv 0$:
 \begin{align}
 -\langle n({\bf x})|\frac{\partial^2{n'({\bf x})}}{\partial {\bf x}^2}\rangle&=-i{\bf \hat p}\cdot{\cal A}_{nn'}({\bf x})+i{\cal A}_{nn'}({\bf x})\cdot{\bf \hat p}+\langle\frac{\partial n({\bf x})}{\partial {\bf x}}|\frac{\partial{n'({\bf x})}}{\partial {\bf x}}\rangle
 \\\nonumber
 \langle\frac{\partial n({\bf x})}{\partial {\bf x}}|\frac{\partial{n'({\bf x})}}{\partial {\bf x}}\rangle&=\sum\limits_{n''}\langle\frac{\partial n({\bf x})}{\partial {\bf x}}| n''({\bf x}) \rangle\langle n''({\bf x})|\frac{\partial{n'({\bf x})}}{\partial {\bf x}}\rangle\equiv -[{\cal A}]^2_{nn'}({\bf x}) \\\nonumber
 \langle n ({\bf x})|[1-i\varepsilon [{\hat K} + {\hat H}^e]\big]|{n' ({\bf x})}\rangle&=\delta_{nn'}-i\varepsilon\left[\frac{[{\bf \hat p}^2\delta_{nn'}-i{\bf \hat p}\cdot {\cal A}_{nn'}({\bf x})- i{\cal A}_{nn'}({\bf x})\cdot{\bf \hat p}-[{\cal A}]^2_{nn'}({\bf x})]}{2m}+H_{nn}({\bf x})\delta_{nn'}\right] \\\nonumber
\end{align}
This leads to the ``Velocity Gauge" molecular Hamiltonian:
 \begin{align}
\hat{H}^{\rm vg} ({\bf x}) ={1\over 2m}{\big[{\bf \hat{p}}-i\hat{{\cal A}}({\bf x})\big]^2}+\hat{E}(\bf x),
\end{align}

\subsection{Equations of Motion for Local Adiabatic Expansion with the ``Length Gauge" molecular Hamiltonian}
Starting from Equation 17 from the main text,
\begin{align}
\dot{c}_n(t)=-\int d{\bf x} ~&g_n^\ast({\bf x}) {\dot g_n({\bf x})} c_n(t) {-}i\sum_{n'} g_n^\ast({\bf x}) H^{\rm la}_{nn'}({\bf x},\bar{\bf x}) {g_{n'}({\bf x})} c_{n'}(t) ~,
\end{align}
and inserting the ``length gauge" adiabatic Hamiltonian, Eq. 5, leads to:
\begin{align}
\dot{c}_n(t)=-\int d{\bf x} ~&g_n^\ast({\bf x}) {\dot g_n({\bf x})} c_n(t) -i g_n^\ast({\bf x})\Big[ {{\bf \hat p}^2\over 2m}  + E_n(\bar{\bf x}) - {\bf f}_n(\bar{\bf x})\cdot({\bf x}-\bar{\bf x})  \Big]g_n^\ast({\bf x})c_n(t)
\\\nonumber
&+\sum_{\substack{n'\neq n}}   g_n^\ast({\bf x}){\cal{\bf {\cal A}}}_{nn^\prime}(\bar{\bf x})\cdot{\bf \bar v}\,e^{i\Delta {\bf p}_{nn^\prime}\cdot({\bf x}-\bar{\bf x})}    {g_{n'}({\bf x})} c_{n'}(t) \,.
\end{align}
We can define a momentum shifted Gaussian, $\tilde g^{n}_{n'}({\bf x}) =e^{i\Delta {\bf p}_{nn^\prime}\cdot(\bar{\bf x}-{\bf x_{n'}})}  e^{i\Delta {\bf p}_{nn^\prime}\cdot({\bf x}-\bar{\bf x})}    {g_{n'}({\bf x})}$, and combine the resulting phase-shift with the real NACV term, ${\cal{\bf {\cal \tilde A}}}_{nn^\prime}(\bar{\bf x},{\bf x}_n')={\cal{\bf {\cal  A}}}_{nn^\prime}(\bar{\bf x})e^{-i\Delta {\bf p}_{nn^\prime}\cdot(\bar{\bf x}-{\bf x}_{n'})}$, which gives:
\begin{align}
\dot{c}_n(t)=-\int d{\bf x} ~&g_n^\ast({\bf x}) {\dot g_n({\bf x})} c_n(t) -i g_n^\ast({\bf x})\Big[ {{\bf \hat p}^2\over 2m}  + E_n(\bar{\bf x}) - {\bf f}_n(\bar{\bf x})\cdot({\bf x}-\bar{\bf x})  \Big]g_n^\ast({\bf x})c_n(t)
\\\nonumber
&-\sum_{\substack{n'\neq n}}  {\cal{\bf {\cal \tilde A}}}_{nn^\prime}(\bar{\bf x},{\bf x}_n')\cdot{\bf \bar v}  \,\int d{\bf x} g_n^\ast({\bf x})\,    {\tilde g^{n}_{n'}({\bf x})} c_{n'}(t)~,
\end{align}
and,
\begin{align}
\dot{c}_n(t)=- i\left(\frac{1}{4}\text{Tr}[{\dot{\alpha}_n^\Re}\cdot\alpha_n^{\Im,-1}]-{\bf p}_n \cdot \dot{{\bf x}}_n\right) c_n(t) &-i \int d{\bf x} g_n^\ast({\bf x})\Big[ {{\bf \hat p}^2\over 2m}  +E_n(\bar{\bf x}) - {\bf f}_n(\bar{\bf x})\cdot({\bf x}-\bar{\bf x})  \Big]g_n^\ast({\bf x})c_n(t)
\\\nonumber
&-\sum_{\substack{n'\neq n}}  {\cal{\bf {\cal \tilde A}}}_{nn^\prime}(\bar{\bf x},{\bf x}_n')\cdot{\bf \bar v}  \,\int d{\bf x} g_n^\ast({\bf x})\,    {\tilde g^{n}_{n'}({\bf x})} c_{n'}(t)~.
\end{align}
Finally, inserting the definition for the $\bf x$ and $\alpha^\Re$ operator equations of motion, Eq. 18 a and c:
\begin{align}
\dot{c}_n(t)&=- i\left(\text{Tr}[{{\alpha}_n^{\Im}\over m}]-{{\bf p}_n^2\over 2m} +E_n(\bar{\bf x}) - {\bf f}_n(\bar{\bf x})\cdot({\bf x}_n-\bar{\bf x})  \right)c_n(t)
\\\nonumber
&-\sum_{\substack{n'\neq n}}  {\cal{\bf {\cal \tilde A}}}_{nn^\prime}(\bar{\bf x},{\bf x}_n')\cdot{\bf \bar v}  \,\int d{\bf x} g_n^\ast({\bf x})\,    {\tilde g^{n}_{n'}({\bf x})} c_{n'}(t)
\\\nonumber
&+\sum_{n'\neq n} \, 2{\rm Im}\{\frac{c_nc^\ast_{n'}}{|c_n|^2} i{\bf {\cal \tilde A}}^\ast_{{n^\prime}n}(\bar{\bf x},{\bf x}_n')\cdot{\bf \bar v} \int d{\bf x} \, \tilde g^{n \ast}_{n'}({\bf x})   [{\bf x}-{\bf x}_{n}]g_n({\bf x})\}\cdot{\bf p}_n
\\\nonumber
&+i\text{Tr}\Big[\frac{1}{2}\text{Re}\Big\{ \frac{c^\ast_nc_{n'}}{|c_n|^2} {i\cal{\bf {\cal \tilde A}}}_{nn^\prime}(\bar{\bf x},{\bf x}_n')\cdot{\bf \bar v} \int d{\bf x} \, g^\ast_n({\bf x})\{4\hat \alpha_n^{\rm Im}\cdot[{\bf x}{-}{\bf x}_n]\otimes  [{\bf x}{-}{\bf x}_n]\cdot \hat \alpha_n^{\rm Im} {-}\hat\alpha_n^{\rm Im} \} \tilde g^{n}_{n'}({\bf x})\Big\}\cdot \alpha_n^{\Im,-1}\Big] ~,
\end{align}
leads to the equation of motion for the coefficents.

\end{document}